# Three-dimensional non-orthogonal multiple-relaxation-time lattice Boltzmann model for multiphase flows

Q. Li[1, *], D. H. Du[1], L. L. Fei[2], Kai H. Luo[3], and Y. Yu[1]

[1]*School of Energy Science and Engineering, Central South University, Changsha 410083, China*

[2]*Departement of Energy and Power Engineering, Tsinghua University, Beijing 100084, China*

[3]*Department of Mechanical Engineering, University College London, Torrington Place, London WC1E 7JE*

**Abstract**

In the classical multiple-relaxation-time (MRT) lattice Boltzmann (LB) method, the transformation matrix is formed by constructing a set of orthogonal basis vectors. In this paper, a theoretical and numerical study is performed to investigate the capability and efficiency of a non-orthogonal MRT-LB model for simulating multiphase flows. First, a three-dimensional non-orthogonal MRT-LB is proposed. A non-orthogonal MRT collision operator is devised based on a set of non-orthogonal basis vectors, through which the transformation matrix and its inverse matrix are considerably simplified as compared with those of an orthogonal MRT collision operator. Furthermore, through the Chapman-Enskog analysis, it is theoretically demonstrated that the three-dimensional non-orthogonal MRT-LB model can correctly recover the macroscopic equations at the Navier-Stokes level in the low Mach number limit. Numerical comparisons between the non-orthogonal MRT-LB model and the usual orthogonal MRT-LB model are made by simulating multiphase flows on the basis of the pseudopotential multiphase LB approach. The numerical results show that, in comparison with the usual orthogonal MRT-LB model, the non-orthogonal MRT-LB model can retain the numerical accuracy while simplifying the implementation.

*Corresponding author: qingli@csu.edu.cn

# I. INTRODUCTION

The lattice Boltzmann (LB) method is becoming an increasingly important numerical approach for a wide range of phenomena and processes [1-8]. This method is based on the mesoscopic kinetic equation for particle distribution function. It simulates fluid flow by tracking the evolution of the particle distribution function, and then the macroscopic averaged properties are obtained by accumulating the distribution function. Compared with the conventional numerical methods, which are based on the direct discretization of macroscopic governing equations, the LB method exhibits some distinctive advantages, such as its inherent parallelizability on multiple processors and easy implementation of fluid-fluid/fluid-solid interactions. In addition, in the conventional numerical methods the convection terms of governing equations are non-linear, while in the LB method the convection terms are linear and the viscous effect is modeled through a linearized collision operator, such as the Bhatnagar-Gross-Krook (BGK) collision operator [2,9], the multiple-relaxation-time (MRT) collision operator [10-16], and the two-relaxation-time (TRT) collision operator [17-20].

Owing to its simplicity, the BGK collision operator is the most frequently used collision operator in the LB community. However, the LB equation using the BGK collision operator is usually found to have stability issues when the viscosity of the fluid is reduced or the Reynolds number is increased. The TRT collision operator is based on the decomposition of the population solution into its symmetric and anti-symmetric components and employs two relaxation parameters to relax the particle distribution function [17,18]. The MRT collision operator is an important extension of the relaxation LB method proposed by Higuera *et al*. [21,22]. The basic idea behind the MRT collision operator is a mapping from the discrete velocity space to the moment space via a transformation matrix $\mathbf{M}$, which allows the moments to be relaxed with individual rates [10-12]. The MRT collision operator has been extensively demonstrated to be capable of improving the numerical stability of LB models by carefully separating the relaxation rates of hydrodynamic and non-hydrodynamic moments [23,24]. The TRT collision operator has certain advantages over the BGK collision operator in terms of numerical stability and accuracy [25] while retaining the simplicity of the BGK collision operator in terms of implementation.

In the literature, the Gram-Schmidt procedure [10,11] is often employed to construct a set of orthogonal basis vectors to form the transformation matrix for an MRT-LB model. This procedure starts with the vectors for the conserved moments (density and momentum). The subsequent step is to take a combination of the velocity vectors $\mathbf{e}_\alpha$ of appropriate order and find the coefficients in such a way that the resulting vector is orthogonal to all the previously found ones [26]. Through the transformation matrix, the particle distribution function can be projected onto the moment space, where the moments are relaxed with individual rates. The relaxed moments are then transformed back to the discrete velocity space and the streaming step of the LB equation is implemented as usual. In most of the existing MRT-LB models, the transformation matrix is an orthogonal matrix. Recently, some research [27-29] showed that the transformation matrix of an MRT-LB model is not necessary to be an orthogonal one. A non-orthogonal transformation matrix for the two-dimensional nine-velocity (D2Q9) lattice can be found in Refs. [27-29]. Moreover, De Rosis [30] showed that a non-orthogonal basis of moments is also efficient in the central-moment-based LB method. Usually, the transformation matrix of a non-orthogonal MRT collision operator is simpler than that of an orthogonal MRT collision operator.

The aim of the present study is to develop a three-dimensional non-orthogonal MRT-LB model and investigate its capability and efficiency for simulating multiphase flows. A non-orthogonal MRT collision operator is devised based on a set of non-orthogonal basis vectors for the three-dimensional nineteen-velocity (D3Q19) lattice. The transformation matrix and its inverse matrix are considerably simplified. The rest of the present paper is organized as follows. The three-dimensional non-orthogonal MRT-LB model is proposed in Section II. Theoretical analysis of the non-orthogonal MRT-LB model is presented in Section III. Numerical investigation is carried out in Section IV and finally a brief summary is given in Section V.

## II. Three-dimensional non-orthogonal MRT-LB model

### A. The MRT-LB framework

In the LB community, the D3Q15 and D3Q19 lattices are the most popular lattice velocity sets for three dimensions [11,12]. The D3Q15 lattice is more computationally efficient than the D3Q19 lattice, while the numerical stability is usually better when using a larger velocity set [11,26]. In the present study, a three-dimensional non-orthogonal MRT-LB model is devised based on the D3Q19 lattice. The model for the D3Q15 lattice can be constructed in a similar way. The MRT-LB equation with a forcing term can be written as follows [12,23]:

$$f_\alpha(\mathbf{x}+\mathbf{e}_\alpha \delta_t, t+\delta_t) = f_\alpha(\mathbf{x},t) - \bar{\Lambda}_{\alpha\beta}\left(f_\beta - f_\beta^{eq}\right)\Big|_{(\mathbf{x},t)} + \frac{\delta_t}{2}\left[G_\alpha\big|_{(\mathbf{x}+\mathbf{e}_\alpha\delta_t, t+\delta_t)} + G_\alpha\big|_{(\mathbf{x},t)}\right], \quad (1)$$

where $f_\alpha$ is the density distribution function, $f_\alpha^{eq}$ is the equilibrium density distribution function, $\mathbf{x}$ is the spatial position, $\mathbf{e}_\alpha$ is the discrete velocity in the $\alpha$ th direction, $t$ is the time, $\delta_t$ is the time step, $G_\alpha$ is the forcing term in the discrete velocity space, and $\bar{\Lambda}_{\alpha\beta} = \left(\mathbf{M}^{-1}\Lambda\mathbf{M}\right)_{\alpha\beta}$ is the collision operator, in which $\mathbf{M}$ is the transformation matrix and $\Lambda$ is a diagonal matrix. The trapezoidal rule has been applied to the forcing term in Eq. (1), which was suggested by He *et al*. [31] in order to achieve second-order accuracy in time.

The lattice velocities $\{\mathbf{e}_\alpha\}$ of the D3Q19 lattice are given by

$$\mathbf{e}_\alpha = \begin{bmatrix} 0 & 1 & -1 & 0 & 0 & 0 & 0 & 1 & -1 & 1 & -1 & 1 & -1 & 1 & -1 & 0 & 0 & 0 & 0 \\ 0 & 0 & 0 & 1 & -1 & 0 & 0 & 1 & -1 & -1 & 1 & 0 & 0 & 0 & 0 & 1 & -1 & 1 & -1 \\ 0 & 0 & 0 & 0 & 0 & 1 & -1 & 0 & 0 & 0 & 0 & 1 & -1 & -1 & 1 & 1 & -1 & -1 & 1 \end{bmatrix}. \quad (2)$$

The implicitness of Eq. (1) can be eliminated by introducing $\bar{f}_\alpha = f_\alpha - 0.5\delta_t G_\alpha$, through which the MRT-LB equation can be transformed to [12,23]:

$$\bar{f}_\alpha(\mathbf{x}+\mathbf{e}_\alpha\delta_t, t+\delta_t) = \bar{f}_\alpha(\mathbf{x},t) - \bar{\Lambda}_{\alpha\beta}\left(\bar{f}_\beta - f_\beta^{eq}\right)\Big|_{(\mathbf{x},t)} + \delta_t\left(G_\alpha - 0.5\bar{\Lambda}_{\alpha\beta}G_\beta\right)\Big|_{(\mathbf{x},t)}. \quad (3)$$

Multiplying Eq. (3) by the transformation matrix $\mathbf{M}$, the right-hand side of Eq. (3), i.e., the collision process, can be implemented in the moment space:

$$\bar{\mathbf{m}}^* = \bar{\mathbf{m}} - \Lambda\left(\bar{\mathbf{m}} - \mathbf{m}^{eq}\right) + \delta_t\left(\mathbf{I} - \frac{\Lambda}{2}\right)\mathbf{S}, \quad (4)$$

where $\mathbf{I}$ is the unit matrix, $\bar{\mathbf{m}} = \mathbf{M}\bar{\mathbf{f}}$, $\mathbf{m}^{eq} = \mathbf{M}\mathbf{f}^{eq}$, and $\mathbf{S} = \mathbf{M}\mathbf{G}$, in which $\bar{\mathbf{f}} = \left(\bar{f}_0, \bar{f}_1, \ldots, \bar{f}_{18}\right)^\mathrm{T}$, $\mathbf{f}^{eq} = \left(f_0^{eq}, f_1^{eq}, \ldots, f_{18}^{eq}\right)^\mathrm{T}$, and $\mathbf{G} = \left(G_0, G_1, \ldots, G_{18}\right)^\mathrm{T}$. Then the streaming process is implemented as follows:

$$\bar{f}_\alpha(\mathbf{x}+\mathbf{e}_\alpha\delta_t, t+\delta_t) = \bar{f}_\alpha^*(\mathbf{x},t), \quad (5)$$

where $\bar{\mathbf{f}}^* = \mathbf{M}^{-1}\bar{\mathbf{m}}^*$ and $\mathbf{M}^{-1}$ is the inverse matrix of the transformation matrix. The macroscopic density and velocity are calculated by

$$\rho = \sum_\alpha \bar{f}_\alpha, \quad \rho\mathbf{u} = \sum_\alpha \mathbf{e}_\alpha \bar{f}_\alpha + \frac{\delta_t}{2}\mathbf{F}, \tag{6}$$

where $\mathbf{F}$ is the total force exerted on the system.

### B. Non-orthogonal MRT-LB model

A three-dimensional non-orthogonal MRT collision operator is now constructed based on the D3Q19 lattice. The following set of non-orthogonal basis vectors is proposed, which can be divided into four groups: (i) the zeroth-order and first-order vectors, which are the vectors related to the conserved moments:

$$M_{0,\alpha} = 1, \quad M_{1,\alpha} = e_{\alpha x}, \quad M_{2,\alpha} = e_{\alpha y}, \quad M_{3,\alpha} = e_{\alpha z}, \tag{7}$$

(ii) the second-order vectors related to the viscous effect at the Navier-Stokes level:

$$M_{4,\alpha} = |\mathbf{e}_\alpha|^2, \quad M_{5,\alpha} = 3e_{\alpha x}^2 - |\mathbf{e}_\alpha|^2, \quad M_{6,\alpha} = e_{\alpha y}^2 - e_{\alpha z}^2,$$

$$M_{7,\alpha} = e_{\alpha x}e_{\alpha y}, \quad M_{8,\alpha} = e_{\alpha x}e_{\alpha z}, \quad M_{9,\alpha} = e_{\alpha y}e_{\alpha z}, \tag{8}$$

(iii) the third-order vectors:

$$M_{10,\alpha} = e_{\alpha x}^2 e_{\alpha y}, \quad M_{11,\alpha} = e_{\alpha x}e_{\alpha y}^2, \quad M_{12,\alpha} = e_{\alpha x}^2 e_{\alpha z},$$

$$M_{13,\alpha} = e_{\alpha x}e_{\alpha z}^2, \quad M_{14,\alpha} = e_{\alpha y}^2 e_{\alpha z}, \quad M_{15,\alpha} = e_{\alpha y}e_{\alpha z}^2, \tag{9}$$

(iv) the fourth-order vectors:

$$M_{16,\alpha} = e_{\alpha x}^2 e_{\alpha y}^2, \quad M_{17,\alpha} = e_{\alpha x}^2 e_{\alpha z}^2, \quad M_{18,\alpha} = e_{\alpha y}^2 e_{\alpha z}^2. \tag{10}$$

The first ten vectors are related to the macroscopic density, momentum, and viscous stress tensor, whereas the additional vectors are related to higher-order moments that do not affect the Navier-Stokes level hydrodynamics. Using such a set of non-orthogonal basis vectors, the relaxation matrix $\Lambda$ (the matrix for relaxation rates) in Eq. (4) can be defined as follows:

$$\Lambda = \mathrm{diag}\left(1, 1, 1, 1, s_e, s_\nu, s_\nu, s_\nu, s_\nu, s_\nu, s_q, s_q, s_q, s_q, s_q, s_q, s_\pi, s_\pi, s_\pi\right), \tag{11}$$

where $s_e$ and $s_\nu$ determine the bulk and shear viscosities, respectively, while $s_q$ and $s_\pi$ are related to non-hydrodynamic moments. The relaxation rates of the conserved moments have been set to 1.0 following Ref. [12]. Note that $M_{4,\alpha}$ in Eq. (8) is related to the energy mode while $M_{5,\alpha}$ and $M_{6,\alpha}$ are retained from the orthogonal MRT-LB method [11,12]. Theoretically, the vectors $M_{4,\alpha}$, $M_{5,\alpha}$, and $M_{6,\alpha}$ can be chosen as $M_{4,\alpha} = e_{\alpha x}^2$, $M_{5,\alpha} = e_{\alpha y}^2$, and $M_{6,\alpha} = e_{\alpha z}^2$, respectively, which will yield a fixed bulk viscosity $\mu_b = 2\mu/3$ when employing a diagonal relaxation matrix like Eq. (11). For such a choice, an alternative approach is to modify the diagonal relaxation matrix as a block-diagonal relaxation matrix to achieve a flexible bulk viscosity [32]. According to Eqs. (7)-(10), the transformation matrix $\mathbf{M}$ is given by

$$\mathbf{M} = \begin{bmatrix}
1 & 1 & 1 & 1 & 1 & 1 & 1 & 1 & 1 & 1 & 1 & 1 & 1 & 1 & 1 & 1 & 1 & 1 & 1 \\
0 & 1 & -1 & 0 & 0 & 0 & 0 & 1 & -1 & 1 & -1 & 1 & -1 & 1 & -1 & 0 & 0 & 0 & 0 \\
0 & 0 & 0 & 1 & -1 & 0 & 0 & 1 & -1 & -1 & 1 & 0 & 0 & 0 & 0 & 1 & -1 & 1 & -1 \\
0 & 0 & 0 & 0 & 0 & 1 & -1 & 0 & 0 & 0 & 0 & 1 & -1 & -1 & 1 & 1 & -1 & -1 & 1 \\
0 & 1 & 1 & 1 & 1 & 1 & 1 & 2 & 2 & 2 & 2 & 2 & 2 & 2 & 2 & 2 & 2 & 2 & 2 \\
0 & 2 & 2 & -1 & -1 & -1 & -1 & 1 & 1 & 1 & 1 & 1 & 1 & 1 & 1 & -2 & -2 & -2 & -2 \\
0 & 0 & 0 & 1 & 1 & -1 & -1 & 1 & 1 & 1 & 1 & -1 & -1 & -1 & -1 & 0 & 0 & 0 & 0 \\
0 & 0 & 0 & 0 & 0 & 0 & 0 & 1 & 1 & -1 & -1 & 0 & 0 & 0 & 0 & 0 & 0 & 0 & 0 \\
0 & 0 & 0 & 0 & 0 & 0 & 0 & 0 & 0 & 0 & 0 & 1 & 1 & -1 & -1 & 0 & 0 & 0 & 0 \\
0 & 0 & 0 & 0 & 0 & 0 & 0 & 0 & 0 & 0 & 0 & 0 & 0 & 0 & 0 & 1 & 1 & -1 & -1 \\
0 & 0 & 0 & 0 & 0 & 0 & 0 & 1 & -1 & -1 & 1 & 0 & 0 & 0 & 0 & 0 & 0 & 0 & 0 \\
0 & 0 & 0 & 0 & 0 & 0 & 0 & 1 & -1 & 1 & -1 & 0 & 0 & 0 & 0 & 0 & 0 & 0 & 0 \\
0 & 0 & 0 & 0 & 0 & 0 & 0 & 0 & 0 & 0 & 0 & 1 & -1 & -1 & 1 & 0 & 0 & 0 & 0 \\
0 & 0 & 0 & 0 & 0 & 0 & 0 & 0 & 0 & 0 & 0 & 1 & -1 & 1 & -1 & 0 & 0 & 0 & 0 \\
0 & 0 & 0 & 0 & 0 & 0 & 0 & 0 & 0 & 0 & 0 & 0 & 0 & 0 & 0 & 1 & -1 & -1 & 1 \\
0 & 0 & 0 & 0 & 0 & 0 & 0 & 0 & 0 & 0 & 0 & 0 & 0 & 0 & 0 & 1 & -1 & 1 & -1 \\
0 & 0 & 0 & 0 & 0 & 0 & 0 & 1 & 1 & 1 & 1 & 0 & 0 & 0 & 0 & 0 & 0 & 0 & 0 \\
0 & 0 & 0 & 0 & 0 & 0 & 0 & 0 & 0 & 0 & 0 & 1 & 1 & 1 & 1 & 0 & 0 & 0 & 0 \\
0 & 0 & 0 & 0 & 0 & 0 & 0 & 0 & 0 & 0 & 0 & 0 & 0 & 0 & 0 & 1 & 1 & 1 & 1
\end{bmatrix}. \quad (12)$$

The inverse matrix of $\mathbf{M}$, namely the matrix $\mathbf{M}^{-1}$, is given in the Appendix A. It can be found that the present non-orthogonal transformation matrix has 145 non-zero elements and its inverse matrix has 96 non-zero elements. However, from Refs. [11,12] we can find that for the D3Q19 lattice the usual transformation matrix and its inverse matrix both have 213 non-zero elements. The matrix-vector calculations $\bar{\mathbf{m}} = \mathbf{M}\bar{\mathbf{f}}$ and $\bar{\mathbf{f}}^* = \mathbf{M}^{-1}\bar{\mathbf{m}}^*$ in Eqs. (4) and (5), respectively, are usually expanded in practical programming [33]. For example, according to the above transformation matrix, the moment $\bar{m}_{18}$ is given by $\bar{m}_{18} = \bar{f}_{15} + \bar{f}_{16} + \bar{f}_{17} + \bar{f}_{18}$. Therefore, reducing the number of non-zero elements in $\mathbf{M}$ and $\mathbf{M}^{-1}$ can simplify the programming and also reduce the computational cost.

According to Eqs. (7)-(10), the equilibria $\mathbf{m}^{eq} = \mathbf{M}\mathbf{f}^{eq}$ in Eq. (4) are given by

$$m_0^{eq} = \rho, \quad m_1^{eq} = \rho u_x, \quad m_2^{eq} = \rho u_y, \quad m_3^{eq} = \rho u_z, \quad m_4^{eq} = \rho + \rho |\mathbf{u}|^2,$$

$$m_5^{eq} = \rho\left(2u_x^2 - u_y^2 - u_z^2\right), \quad m_6^{eq} = \rho\left(u_y^2 - u_z^2\right), \quad m_7^{eq} = \rho u_x u_y, \quad m_8^{eq} = \rho u_x u_z, \quad m_9^{eq} = \rho u_y u_z,$$

$$m_{10}^{eq} = \rho c_s^2 u_y, \quad m_{11}^{eq} = \rho c_s^2 u_x, \quad m_{12}^{eq} = \rho c_s^2 u_z, \quad m_{13}^{eq} = \rho c_s^2 u_x, \quad m_{14}^{eq} = \rho c_s^2 u_z, \quad m_{15}^{eq} = \rho c_s^2 u_y,$$

$$m_{16}^{eq} = \varphi + \rho c_s^2 \left(u_x^2 + u_y^2\right), \quad m_{17}^{eq} = \varphi + \rho c_s^2 \left(u_x^2 + u_z^2\right), \quad m_{18}^{eq} = \varphi + \rho c_s^2 \left(u_y^2 + u_z^2\right), \tag{13}$$

where $c_s^2 = 1/3$ and $\varphi = \rho c_s^4 \left(1 - 1.5|\mathbf{u}|^2\right)$. Correspondingly, the forcing term $\mathbf{S}$ in Eq. (4) is given by

$$\mathbf{S} = \begin{bmatrix} 0 \\ F_x \\ F_y \\ F_z \\ 2\mathbf{F}\cdot\mathbf{u} \\ 2\left(2F_x u_x - F_y u_y - F_z u_z\right) \\ 2\left(F_y u_y - F_z u_z\right) \\ F_x u_y + F_y u_x \\ F_x u_z + F_z u_x \\ F_y u_z + F_z u_y \\ c_s^2 F_y \\ c_s^2 F_x \\ c_s^2 F_z \\ c_s^2 F_x \\ c_s^2 F_z \\ c_s^2 F_y \\ 2c_s^2 \left(u_x F_x + u_y F_y\right) \\ 2c_s^2 \left(u_x F_x + u_z F_z\right) \\ 2c_s^2 \left(u_y F_y + u_z F_z\right) \end{bmatrix}, \tag{14}$$

where $\mathbf{F}$ is the total force exerted on the system. In the pseudopotential multiphase LB approach, the pseudopotential interaction force is given by [6]:

$$\mathbf{F}_m = -G\psi(\mathbf{x}) \sum_\alpha w_\alpha \psi\left(\mathbf{x} + \mathbf{e}_\alpha \delta_t\right) \mathbf{e}_\alpha, \tag{15}$$

where $G$ is the interaction strength, $\psi(\mathbf{x})$ is the pseudopotential, and $w_\alpha$ are the weights. For the D3Q19 lattice, the weights $w_\alpha$ in Eq. (15) are given by $w_{1-6} = 1/6$ and $w_{7-18} = 1/12$. In the literature, two types of pseudopotentials are widely used. One is the exponential-form pseudopotential [34], i.e., $\psi(\mathbf{x}) = \psi_0 \exp(-\rho_0/\rho)$, where $\psi_0$ and $\rho_0$ are constant, and the other is the square-root-form pseudopotential $\psi(\mathbf{x}) = \sqrt{2(p_{EOS} - \rho c_s^2)/Gc^2}$ [35,36], in which $c=1$ is the lattice constant and $p_{EOS}$ is a prescribed non-ideal equation of state.

Using the square-root-form pseudopotential, the pseudopotential multiphase LB model usually suffers from the problem of thermodynamic inconsistency [37], namely the coexistence curve predicted by the pseudopotential LB model is inconsistent with that given by the Maxwell equal-area law. To solve this problem, Li *et al*. [38,39] proposed that the thermodynamic consistency can be achieved by adjusting the mechanical stability condition of the pseudopotential LB model through an improved forcing scheme. Moreover, in the original pseudopotential LB model the surface tension is dependent on the density ratio. An alternative approach has been developed by Li and Luo [40] to decouple the surface tension from the density ratio. Some extensions of Li *et al*.'s approaches [38-40] have been made by Zhang *et al*. [41], Xu *et al*. [42], Lycett-Brown and Luo [43], and Ammar *et al*. [44]. To achieve thermodynamic consistency, the fifth moment of the forcing term $\mathbf{S}$ in Eq. (14) can be changed to

$$S_4 = 2\mathbf{F} \cdot \mathbf{u} + \frac{6\sigma |\mathbf{F}_m|^2}{\psi^2 \delta_t (s_e^{-1} - 0.5)}, \qquad (16)$$

where the constant $\sigma$ is employed to adjust the mechanical stability condition of the pseudopotential LB model [39,41,42].

## III. Theoretical analysis

In this section, the Chapman-Enskog analysis is performed for the three-dimensional non-orthogonal MRT-LB model, which can be implemented by introducing the following multi-scale expansions:

$$\partial_t = \partial_{t0} + \delta_t \partial_{t1}, \quad f_\alpha = f_\alpha^{eq} + \delta_t f_\alpha^{(1)} + \delta_t^2 f_\alpha^{(2)}. \qquad (17)$$

The Taylor series expansion of Eq. (1) yields

$$\delta_t (\partial_t + \mathbf{e}_\alpha \cdot \nabla) f_\alpha + \frac{\delta_t^2}{2} (\partial_t + \mathbf{e}_\alpha \cdot \nabla)^2 f_\alpha = -\bar{\Lambda}_{\alpha\beta} (f_\beta - f_\beta^{eq})|_{(\mathbf{x},t)} + \delta_t G_\alpha + \frac{\delta_t^2}{2} (\partial_t + \mathbf{e}_\alpha \cdot \nabla) G_\alpha. \qquad (18)$$

Using the multi-scale expansions, Eq. (18) can be rewritten in the consecutive orders of $\delta_t$ as follows:

$$O(\delta_t): \quad (\partial_{t0} + \mathbf{e}_\alpha \cdot \nabla) f_\alpha^{eq} = -\bar{\Lambda}_{\alpha\beta} f_\beta^{(1)} |_{(\mathbf{x},t)} + G_\alpha, \tag{19}$$

$$O(\delta_t^2): \quad \partial_{t1} f_\alpha^{eq} + (\partial_{t0} + \mathbf{e}_\alpha \cdot \nabla) f_\alpha^{(1)} + \frac{1}{2}(\partial_{t0} + \mathbf{e}_\alpha \cdot \nabla)^2 f_\alpha^{eq} = -\bar{\Lambda}_{\alpha\beta} f_\beta^{(2)} |_{(\mathbf{x},t)} + \frac{1}{2}(\partial_t + \mathbf{e}_\alpha \cdot \nabla) G_\alpha. \tag{20}$$

Multiplying Eqs. (19) and (20) by the transformation matrix $\mathbf{M}$ leads to the following equations:

$$O(\delta_t): \quad \mathbf{D}_0 \mathbf{m}^{eq} = -\Lambda \mathbf{m}^{(1)} + \mathbf{S}, \tag{21}$$

$$O(\delta_t^2): \quad \partial_{t1} \mathbf{m}^{eq} + \mathbf{D}_0 \mathbf{m}^{(1)} + \frac{1}{2} \mathbf{D}_0^2 \mathbf{m}^{eq} = -\Lambda \mathbf{m}^{(2)} + \frac{1}{2} \mathbf{D}_0 \mathbf{S}, \tag{22}$$

where $\mathbf{D}_0 = \partial_{t0} \mathbf{I} + \mathbf{C}_i \partial_i$, in which $\mathbf{I}$ is the unit matrix, $\mathbf{C}_i \partial_i = \mathbf{C}_x \partial_x + \mathbf{C}_y \partial_y + \mathbf{C}_z \partial_z$, and $\mathbf{C}_i = \mathbf{M} \mathbf{E}_i \mathbf{M}^{-1}$ with $\mathbf{E}_i = \text{diag}(\mathbf{e}_{0,i}, \mathbf{e}_{1,i}, ..., \mathbf{e}_{18,i})$. Substituting Eq. (21) into Eq. (22), we can obtain

$$\partial_{t1} \mathbf{m}^{eq} + \mathbf{D}_0 \left( \mathbf{I} - \frac{\Lambda}{2} \right) \mathbf{m}^{(1)} = -\Lambda \mathbf{m}^{(2)}. \tag{23}$$

The matrix $\mathbf{C}_i$ in $\mathbf{D}_0$ can be obtained according to the lattice velocities $\{\mathbf{e}_\alpha\}$, the transformation matrix $\mathbf{M}$, and its inverse matrix $\mathbf{M}^{-1}$. According to the expression of $\mathbf{C}_i$ and Eq. (21), we can obtain

$$\partial_{t0} \rho + \partial_x m_1^{eq} + \partial_y m_2^{eq} + \partial_z m_3^{eq} = 0, \tag{24}$$

$$\partial_{t0}(\rho u_x) + \frac{1}{3} \partial_x \left( m_4^{eq} + m_5^{eq} \right) + \partial_y m_7^{eq} + \partial_z m_8^{eq} = F_x, \tag{25}$$

$$\partial_{t0}(\rho u_y) + \partial_x m_7^{eq} + \partial_y \left( \frac{1}{3} m_4^{eq} - \frac{1}{6} m_5^{eq} + \frac{1}{2} m_6^{eq} \right) + \partial_z m_9^{eq} = F_y, \tag{26}$$

$$\partial_{t0}(\rho u_z) + \partial_x m_8^{eq} + \partial_y m_9^{eq} + \partial_z \left( \frac{1}{3} m_4^{eq} - \frac{1}{6} m_5^{eq} - \frac{1}{2} m_6^{eq} \right) = F_z. \tag{27}$$

Substituting Eq. (13) into the above equations leads to

$$\partial_{t0} \rho + \partial_x (\rho u_x) + \partial_y (\rho u_y) + \partial_z (\rho u_z) = 0, \tag{28}$$

$$\partial_{t0}(\rho u_x) + \partial_x (p + \rho u_x^2) + \partial_y (\rho u_y u_x) + \partial_z (\rho u_z u_x) = F_x, \tag{29}$$

$$\partial_{t0}(\rho u_y) + \partial_x (\rho u_x u_y) + \partial_y (p + \rho u_y^2) + \partial_z (\rho u_z u_y) = F_y, \tag{30}$$

$$\partial_{t0}(\rho u_z) + \partial_x (\rho u_x u_z) + \partial_y (\rho u_y u_z) + \partial_z (p + \rho u_z^2) = F_z, \tag{31}$$

where $p = \rho c_s^2$. Similarly, from Eq. (23) we can obtain

$$\partial_{t1} \rho = 0, \tag{32}$$

$$\partial_{t1}(\rho u_x) + \frac{1}{3} \partial_x \left[ \left(1 - \frac{s_e}{2}\right) m_4^{(1)} + \left(1 - \frac{s_\nu}{2}\right) m_5^{(1)} \right] + \left(1 - \frac{s_\nu}{2}\right) \partial_y m_7^{(1)} + \left(1 - \frac{s_\nu}{2}\right) \partial_z m_8^{(1)} = 0, \tag{33}$$

$$\partial_{t1}(\rho u_y) + \left(1-\frac{s_v}{2}\right)\partial_x m_7^{(1)} + \partial_y\left[\frac{1}{3}\left(1-\frac{s_e}{2}\right)m_4^{(1)} - \frac{1}{6}\left(1-\frac{s_v}{2}\right)m_5^{(1)} + \frac{1}{2}\left(1-\frac{s_v}{2}\right)m_6^{(1)}\right] + \left(1-\frac{s_v}{2}\right)\partial_z m_9^{(1)} = 0, \quad (34)$$

$$\partial_{t1}(\rho u_z) + \left(1-\frac{s_v}{2}\right)\partial_x m_8^{(1)} + \left(1-\frac{s_v}{2}\right)\partial_y m_9^{(1)} + \partial_z\left[\frac{1}{3}\left(1-\frac{s_e}{2}\right)m_4^{(1)} - \frac{1}{6}\left(1-\frac{s_v}{2}\right)m_5^{(1)} - \frac{1}{2}\left(1-\frac{s_v}{2}\right)m_6^{(1)}\right] = 0. \quad (35)$$

Meanwhile, according to Eq. (21) we have

$$\partial_{t0}m_4^{eq} + \partial_x\left(m_1^{eq} + m_{11}^{eq} + m_{13}^{eq}\right) + \partial_y\left(m_2^{eq} + m_{10}^{eq} + m_{15}^{eq}\right) + \partial_z\left(m_3^{eq} + m_{12}^{eq} + m_{14}^{eq}\right) = -s_e m_4^{(1)} + S_4, \quad (36)$$

$$\partial_{t0}m_5^{eq} + \partial_x\left(2m_1^{eq} - m_{11}^{eq} - m_{13}^{eq}\right) + \partial_y\left(-m_2^{eq} + 2m_{10}^{eq} - m_{15}^{eq}\right) + \partial_z\left(-m_3^{eq} + 2m_{12}^{eq} - m_{14}^{eq}\right) = -s_v m_5^{(1)} + S_5, \quad (37)$$

$$\partial_{t0}m_6^{eq} + \partial_x\left(m_{11}^{eq} - m_{13}^{eq}\right) + \partial_y\left(m_2^{eq} - m_{15}^{eq}\right) + \partial_z\left(-m_3^{eq} + m_{14}^{eq}\right) = -s_v m_6^{(1)} + S_6, \quad (38)$$

$$\partial_{t0}m_7^{eq} + \partial_x m_{10}^{eq} + \partial_y m_{11}^{eq} = -s_v m_7^{(1)} + S_7, \quad (39)$$

$$\partial_{t0}m_8^{eq} + \partial_x m_{12}^{eq} + \partial_z m_{13}^{eq} = -s_v m_8^{(1)} + S_8, \quad (40)$$

$$\partial_{t0}m_9^{eq} + \partial_y m_{14}^{eq} + \partial_z m_{15}^{eq} = -s_v m_9^{(1)} + S_9. \quad (41)$$

Substituting Eqs. (13) and (14) into Eqs. (36)-(41) yields

$$\partial_{t0}\left(\rho + \rho|\mathbf{u}|^2\right) + \partial_x\left(\rho u_x + 2\rho c_s^2 u_x\right) + \partial_y\left(\rho u_y + 2\rho c_s^2 u_y\right) + \partial_z\left(\rho u_z + 2\rho c_s^2 u_z\right) = -s_e m_4^{(1)} + 2\mathbf{F}\cdot\mathbf{u}, \quad (42)$$

$$\partial_{t0}\left[\rho\left(2u_x^2 - u_y^2 - u_z^2\right)\right] + 4\partial_x\left(\rho c_s^2 u_x\right) - 2\partial_y\left(\rho c_s^2 u_y\right) - 2\partial_z\left(\rho c_s^2 u_z\right) = -s_v m_5^{(1)} + 2\left(2F_x u_x - F_y u_y - F_z u_z\right), \quad (43)$$

$$\partial_{t0}\left[\rho\left(u_y^2 - u_z^2\right)\right] + 2\partial_y\left(\rho c_s^2 u_y\right) - 2\partial_z\left(\rho c_s^2 u_z\right) = -s_v m_6^{(1)} + 2\left(F_y u_y - F_z u_z\right), \quad (44)$$

$$\partial_{t0}(\rho u_x u_y) + \partial_x\left(\rho c_s^2 u_x\right) + \partial_y\left(\rho c_s^2 u_y\right) = -s_v m_7^{(1)} + F_x u_y + F_y u_x, \quad (45)$$

$$\partial_{t0}(\rho u_x u_z) + \partial_x\left(\rho c_s^2 u_x\right) + \partial_z\left(\rho c_s^2 u_z\right) = -s_v m_8^{(1)} + F_x u_z + F_z u_x, \quad (46)$$

$$\partial_{t0}(\rho u_y u_z) + \partial_y\left(\rho c_s^2 u_y\right) + \partial_z\left(\rho c_s^2 u_z\right) = -s_v m_9^{(1)} + F_y u_z + F_z u_y. \quad (47)$$

The equilibrium moments $m_5^{eq} = \rho\left(2u_x^2 - u_y^2 - u_z^2\right)$, $m_6^{eq} = \rho\left(u_y^2 - u_z^2\right)$, $m_7^{eq} = \rho u_x u_y$, $m_8^{eq} = \rho u_x u_z$, and $m_9^{eq} = \rho u_y u_z$ can also be found in the classical orthogonal MRT-LB models [12]. It can be readily verified that Eqs. (43)-(47) are consistent with those obtained from the orthogonal MRT-LB models [12]. The difference lies in the form of Eq. (42). In the present non-orthogonal MRT-LB model, the equilibrium moment related to the energy mode is given by $e^{eq} = m_4^{eq} = \rho + \rho|\mathbf{u}|^2$, while in the orthogonal MRT-LB model $e^{eq} = -\rho + \rho|\mathbf{u}|^2$ (taking the D3Q15 model for example) and the equation of $e^{eq}$ at $t_0$ time scale is given by (see Eq. (45) in Ref. [12])

$$\partial_{t0}\left(-\rho+\rho|\mathbf{u}|^2\right)-\frac{1}{3}\partial_x(\rho u_x)-\frac{1}{3}\partial_y(\rho u_y)-\frac{1}{3}\partial_z(\rho u_z)=-s_e e^{(1)}+2\mathbf{F}\cdot\mathbf{u}. \tag{48}$$

By combining Eq. (42) with Eq. (28), we can rewrite Eq. (42) as follows:

$$\partial_{t0}\left(-\rho+\rho|\mathbf{u}|^2\right)+\partial_x\left(-\rho u_x+2\rho c_s^2 u_x\right)+\partial_y\left(-\rho u_y+2\rho c_s^2 u_y\right)+\partial_z\left(-\rho u_z+2\rho c_s^2 u_z\right)=-s_e m_4^{(1)}+2\mathbf{F}\cdot\mathbf{u}. \tag{49}$$

Since $c_s^2=1/3$, it can be found that Eq. (49) and Eq. (48) are identical.

With the help of Eqs. (28)-(31), we can derive the following equations from Eqs. (42)-(47) (the third-order velocity terms are neglected according to the low Mach number limit):

$$-s_e m_4^{(1)} \approx 2\rho c_s^2\left(\partial_x u_x+\partial_y u_y+\partial_z u_z\right), \quad -s_v m_5^{(1)} \approx 2\rho c_s^2\left(2\partial_x u_x-\partial_y u_y-\partial_z u_z\right),$$

$$-s_v m_6^{(1)} \approx 2\rho c_s^2\left(\partial_y u_y-\partial_z u_z\right), \quad -s_v m_7^{(1)} \approx \rho c_s^2\left(\partial_x u_y+\partial_y u_x\right),$$

$$-s_v m_8^{(1)} \approx \rho c_s^2\left(\partial_x u_z+\partial_z u_x\right), \quad -s_v m_9^{(1)} \approx \rho c_s^2\left(\partial_y u_z+\partial_z u_y\right). \tag{50}$$

Substituting Eq. (50) into Eqs. (33)-(35) and then multiplying the results with $\delta_t$, we can obtain

$$\delta_t\partial_{t1}(\rho u_x)=\partial_x\left[\mu_b(\nabla\cdot\mathbf{u})+\frac{2\mu}{3}\left(2\partial_x u_x-\partial_y u_y-\partial_z u_z\right)\right]+\partial_y\left[\mu\left(\partial_y u_x+\partial_x u_y\right)\right]+\partial_z\left[\mu\left(\partial_z u_x+\partial_x u_z\right)\right]=0, \tag{51}$$

$$\delta_t\partial_{t1}(\rho u_y)=\partial_x\left[\mu\left(\partial_x u_y+\partial_y u_x\right)\right]+\partial_y\left[\mu_b(\nabla\cdot\mathbf{u})+\frac{2\mu}{3}\left(2\partial_y u_y-\partial_x u_x-\partial_z u_z\right)\right]+\partial_z\left[\mu\left(\partial_z u_y+\partial_y u_z\right)\right]=0, \tag{52}$$

$$\delta_t\partial_{t1}(\rho u_z)=\partial_x\left[\mu\left(\partial_x u_z+\partial_z u_x\right)\right]+\partial_y\left[\mu\left(\partial_y u_z+\partial_z u_y\right)\right]+\partial_z\left[\mu_b(\nabla\cdot\mathbf{u})+\frac{2\mu}{3}\left(2\partial_z u_z-\partial_x u_x-\partial_y u_y\right)\right]=0, \tag{53}$$

where the dynamic shear viscosity $\mu$ and the bulk viscosity $\mu_b$ are given by

$$\mu=\rho c_s^2\left(\frac{1}{s_v}-\frac{1}{2}\right)\delta_t, \quad \mu_b=\frac{2}{3}\rho c_s^2\left(\frac{1}{s_e}-\frac{1}{2}\right)\delta_t. \tag{54}$$

Combining Eq. (28) with Eq. (32) through $\partial_t=\partial_{t0}+\delta_t\partial_{t1}$, the continuity equation can be obtained

$$\partial_t\rho+\nabla\cdot(\rho\mathbf{u})=0, \tag{55}$$

Similarly, combining Eqs. (29)-(31) with Eqs. (51)-(53), we can obtain the Navier-Stokes equation as follows:

$$\partial_t(\rho\mathbf{u})+\nabla\cdot(\rho\mathbf{uu})=-\nabla p+\nabla\cdot\left[\mu\left(\nabla\mathbf{u}+(\nabla\mathbf{u})^T\right)-\frac{2}{3}\mu(\nabla\cdot\mathbf{u})\mathbf{I}+\mu_b(\nabla\cdot\mathbf{u})\mathbf{I}\right]+\mathbf{F}. \tag{56}$$

The above analysis clearly shows that the macroscopic equations at the Navier-Stokes level can be correctly recovered from the three-dimensional non-orthogonal MRT-LB model in the low Mach number limit. Note that, when the square-root-form pseudopotential is employed, the fifth moment of the forcing term is given by Eq. (16), which will introduce an additional term into the Navier-Stokes equation to modify the pressure tensor and adjust the mechanical stability condition of the pseudopotential LB model [39,41,42].

## IV. Numerical simulations

In this section, numerical simulations are carried out to investigate the capability and efficiency of the three dimensional non-orthogonal MRT-LB model for simulating multiphase flows. In particular, comparisons between the non-orthogonal MRT-LB model and the usual orthogonal MRT-LB model will be made so as to identify whether the non-orthogonal MRT-LB model can serve as an alternative to the usual orthogonal MRT-LB model. The exponential-form pseudopotential is employed in Sec. A to Sec. C and the simulations using the square-root-form pseudopotential are performed in the last two subsections.

### A. Phase separation

First, we consider three-dimensional phase separation in a cubic domain of $120 \times 120 \times 120$ with periodic boundary conditions in all directions. The exponential-form pseudopotential $\psi(\mathbf{x}) = \psi_0 \exp(-\rho_0/\rho)$ is adopted. For such a pseudopotential, the thermodynamic consistency or the Maxwell equal-area law is satisfied as long as the macroscopic equations at the Navier-Stokes level are correctly recovered. Some previous studies [45,46] have shown that the numerical coexistence densities produced by the pseudopotential LB model are very sensitive to the error terms in the recovered macroscopic equations. In the present study, we adopt $\psi_0 = 1$, $\rho_0 = 1$, and $G = -10/3$ [46], which leads to the coexistence densities $\rho_L \approx 2.783$ and $\rho_V \approx 0.3675$ according to the Maxwell equal-area law.

Initially, the density in the computational domain is taken as $\rho = \rho_0 - N_{\text{rand}}/5000$, where $N_{\text{rand}}$ is a random number in the interval $[0, 10]$. The relaxation parameter $s_e$, which determines the bulk viscosity, is chosen as $s_e = 0.8$, the relaxation parameter $s_\nu$ changes with the kinematic viscosity $\nu = \mu/\rho$, which varies from $\nu = 0.01$ to $0.15$, and the other relaxation parameters are fixed at $1.2$. The same choices are applied to the usual orthogonal MRT-LB model based on the D3Q19 lattice [11,12]. Some snapshots of the results obtained by the non-orthogonal MRT-LB model with $v = 0.1$ are displayed in Fig. 1. During the phase separation process, the system changes from single phase to two phases. The equilibrium state of the system can be observed in Fig. 1(d), which is taken at $t = 8000\delta_t$. From Fig. 1(d) we can see that the region occupied by the liquid is in the form of a cylinder with the liquid density $\rho_L \approx 2.801$ inside the cylinder and the vapor density $\rho_V \approx 0.369$ outside the cylinder, which are in good agreement with the coexistence liquid and vapor densities given by the Maxwell equal-area law.

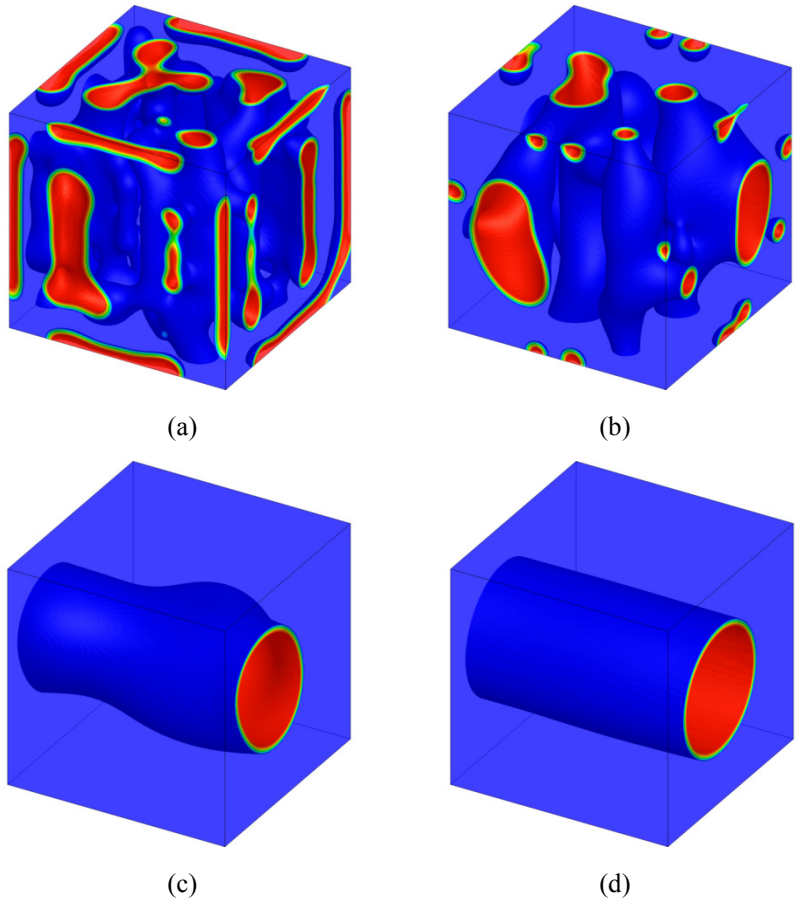

**Fig. 1** Snapshots of three-dimensional phase separation at (a) $t = 400\delta_t$, (b) $800\delta_t$, (c) $2800\delta_t$, and (d) $8000\delta_t$. The pseudopotential is taken as $\psi(\mathbf{x}) = \exp(-1/\rho)$ and the interaction strength is chosen as $G = -10/3$. The coexistence liquid and vapor densities obtained from the Maxwell equal-area law are given by $\rho_L \approx 2.783$ and $\rho_V \approx 0.3675$, respectively.

Table I depicts a comparison of the numerical coexistence densities obtained by the non-orthogonal

MRT-LB model and the orthogonal MRT-LB model when the kinematic viscosity varies from $v = 0.01$ to $0.15$. For both models, it can be seen that their numerical results agree well with the results given by the Maxwell equal-area law. More importantly, in all the cases there are only very minor differences between the results of the non-orthogonal MRT-LB model and those of the orthogonal MRT-LB model. Previously, it has been mentioned that the pseudopotential LB model is very sensitive to the additional (error) terms in the recovered macroscopic equations. The good agreement shown in Table I numerically confirms that the non-orthogonal MRT-LB model can recover the correct macroscopic equations at the Navier-Stokes level. Furthermore, we find that in this test both models are unstable when the kinematic viscosity is taken as $v = 10^{-3}$, but stable with $v = 2 \times 10^{-3}$, which implies the numerical stability of the MRT-LB method is retained when employing a non-orthogonal MRT collision operator to simplify the implementation.

**Table I**. Comparison of the coexistence densities obtained by the non-orthogonal and orthogonal MRT-LB models.

| Model | $v = 0.01$ | | $v = 0.05$ | | $v = 0.1$ | | $v = 0.15$ | |
|---|---|---|---|---|---|---|---|---|
| | $\rho_L$ | $\rho_V$ | $\rho_L$ | $\rho_V$ | $\rho_L$ | $\rho_V$ | $\rho_L$ | $\rho_V$ |
| Non-orthogonal | 2.794 | 0.365 | 2.794 | 0.368 | 2.801 | 0.369 | 2.797 | 0.369 |
| Orthogonal | 2.792 | 0.366 | 2.791 | 0.367 | 2.797 | 0.369 | 2.798 | 0.369 |

### B. Static droplets

In this subsection, the Laplace law for a static droplet immersed in its vapor phase is employed to examine the three-dimensional non-orthogonal MRT-LB model. In three-dimensional space, the Laplace law is given by $\Delta p = p_{\text{in}} - p_{\text{out}} = 2\vartheta/R_{\text{d}}$, where $p_{\text{in}}$ and $p_{\text{out}}$ are the fluid pressures inside and outside the droplet, respectively, $R_{\text{d}}$ is the droplet radius, and $\vartheta$ is the surface tension. When the surface tension is given, the pressure difference is proportional to $1/R_{\text{d}}$. Simulations are carried out in a cubic domain of $150 \times 150 \times 150$ with periodic boundary conditions in all directions. The kinematic viscosity is taken as $v = 0.1$, which corresponds to $s_v^{-1} = 0.8$. The other relaxation parameters are the same as those used in the previous subsection. A spherical droplet is initially located at the center of the computational domain. The pressure difference is measured at $t = 2 \times 10^4 \delta_t$, at which the equilibrium state is approximately achieved. The numerical results are plotted in Fig. 2. For comparison, the results of the orthogonal MRT-LB model are also shown in the figure. The linear relationship between the pressure difference and $1/R_d$ can be clearly observed for both models. Similar to the previous test, the present test also shows that there are only very minor differences between the results of the non-orthogonal MRT-LB model and the orthogonal MRT-LB model.

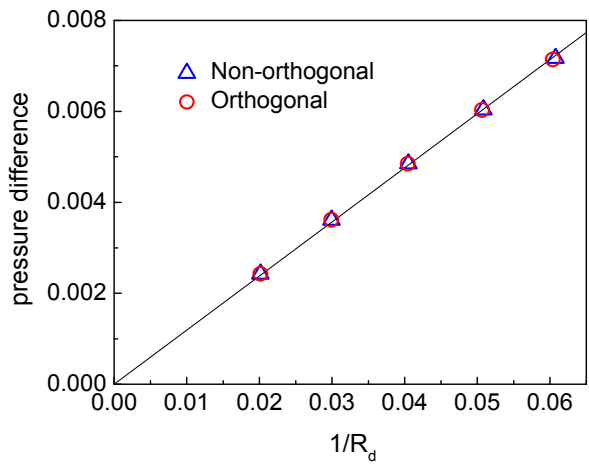

**Fig. 2** Validation of the Laplace law.

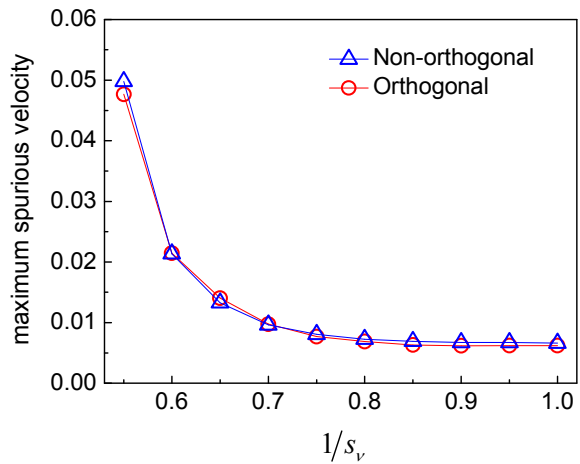

**Fig. 3** Comparison of the maximum spurious velocities given by the non-orthogonal MRT-LB model and the orthogonal MRT-LB model.

Furthermore, we also compare the maximum spurious velocities yielded by the non-orthogonal MRT-LB model and the orthogonal MRT-LB model. The spurious velocities, also called spurious currents, have been observed in almost all the simulations of multiphase flows involving curved interfaces. The grid system is now chosen as $N_x \times N_y \times N_z = 80 \times 80 \times 80$ and the droplet radius is fixed at $15$. The relaxation parameter $s_\nu$ varies from $s_\nu^{-1} = 0.55$ to $1.0$. The maximum spurious velocities produced by the non-orthogonal and orthogonal MRT-LB models are compared in Fig. 3, from which no significant differences are observed between the results of the two models. In addition, for both models it can be found that the maximum spurious velocity increases significantly when $s_\nu^{-1}$ is close to $0.5$. Such a phenomenon is consistent with the findings in the literature for other multiphase LB models. For the pseudopotential LB model, the spurious velocities can be reduced by using high-order isotropic gradient operators to calculate the interaction force or widening the interface [6].

### C. Contact angles

The capability of the three-dimensional non-orthogonal MRT-LB model for simulating contact angles is examined in this subsection. There have been many studies of wetting phenomena using the pseudopotential multiphase LB method [6] and applying a fluid-solid interaction to implement contact angles is the most frequently used treatment in the pseudopotential LB method, which was introduced by Martys and Chen in 1996 [47]. Since then different types of fluid-solid interactions have been developed, which have been reviewed in Ref. [48]. In recent years, the geometric formulation [49,50], originally devised for implementing contact angles in the phase-field method, has also been applied to the pseudopotential LB method [51,52]. In three-dimensional space, the geometric formulation for the pseudopotential LB method can be given by [52]

$$\rho_{i,j,0} = \rho_{i,j,2} + \phi \tan\left(90^\circ - \theta_a\right), \tag{57}$$

where $\phi = \sqrt{\left(\rho_{i+1,j,1} - \rho_{i-1,j,1}\right)^2 + \left(\rho_{i,j+1,1} - \rho_{i,j-1,1}\right)^2}$, $\theta_a$ is an analytically prescribed contact angle, and $\rho_{i,j,0}$ represents the density at the ghost layer $(i, j, 0)$ beneath the solid wall. The first and the second indexes represent the coordinates along the *x*- and *y*-directions, respectively, while the third index denotes the coordinate normal to the solid wall.

In our simulations, the grid system is taken as $N_x \times N_y \times N_z = 160 \times 160 \times 120$. The kinematic viscosity is set to $\nu = 0.1$. Initially, a spherical droplet of radius $r_0 = 30$ is placed on the bottom surface. The non-slip boundary condition is applied at the solid surfaces and the periodic boundary condition is utilized in the *x* and *y* directions. The modeling results are displayed in Fig. 4 with the prescribed contact angle $\theta_a$ in Eq. (57) being setting to $30^\circ$, $60^\circ$, $90^\circ$, and $135^\circ$, respectively. According to the numerical results, the contact angles produced by the non-orthogonal MRT-LB model are $\theta \approx 29.1^\circ$, $60.8^\circ$, $89.93^\circ$, and $135.4^\circ$, respectively, which are in good agreement with the analytically prescribed contact angles and the maximum error is within $\pm 1^\circ$, which demonstrates the capability of the non-orthogonal MRT-LB model for simulating contact angles.

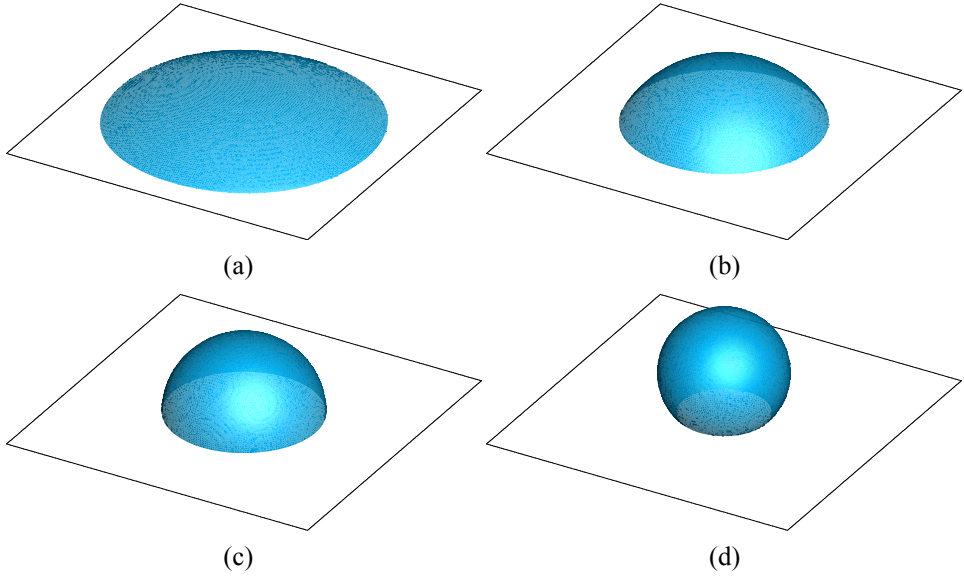

**Fig. 4** Simulation of contact angles using the non-orthogonal MRT-LB model. (a) $\theta \approx 29.1^\circ$, (b) $\theta \approx 60.8^\circ$, (c) $\theta \approx 89.93^\circ$, and (d) $\theta \approx 135.4^\circ$.

### D. The square-root-form pseudopotential

In this subsection, some simulations are performed using the square-root-form pseudopotential. A comprehensive review of the pseudopotential multiphase LB method using this type of pseudopotentials can be found in Ref. [6]. Here the Carnahan-Starling (C-S) equation of state is adopted [36]

$$p_{\text{EOS}} = \rho RT \frac{1 + b\rho/4 + (b\rho/4)^2 - (b\rho/4)^3}{(1 - b\rho/4)^3} - a\rho^2, \tag{58}$$

where $a = 0.4963R^2 T_c^2/p_c$ and $b = 0.18727RT_c/p_c$ with $T_c = 0.37733a/(bR)$. Using the square-root-form pseudopotential, the only requirement for $G$ is to ensure that the whole term inside the square root is positive. The parameters $R$ and $b$ are taken as $R=1$ and $b=4$, respectively. But note that the interface thickness can be adjusted by tuning the parameter $a$ in the equation of state [39].

First, the numerical coexistence curve predicted by the non-orthogonal MRT-LB model is compared with the analytical coexistence curve given by the Maxwell equal-area law through simulating flat interfaces. The grid system is taken as $N_x \times N_y \times N_z = 100 \times 100 \times 100$ with periodic boundary conditions in all directions. The flat liquid-vapor interfaces are located at $z = 0.25N_z$ and $z = 0.75N_z$. For simplicity, the parameter $a$ in the C-S equation of state is chosen as $a = 0.5$ for all the investigated reduced temperatures, but it should be noted that the interface thickness usually decreases with the decrease of the reduced temperature (see Fig. 3 in Ref. [39] for details). The constant $\sigma$ in Eq. (16) is set to $0.116$ for flat interfaces. The parameter $s_\nu$ is chosen as $s_\nu^{-1} = 0.8$, which corresponds to the kinematic viscosity $\nu = 0.1$, and the other parameters are the same as those used in the previous subsections. The numerical results are displayed in Fig. 5, from which we can see that the numerical coexistence curve produced by the non-orthogonal MRT-LB model agrees well with the analytical coexistence curve given by the Maxwell equal-area law in a wide range of reduced temperatures.

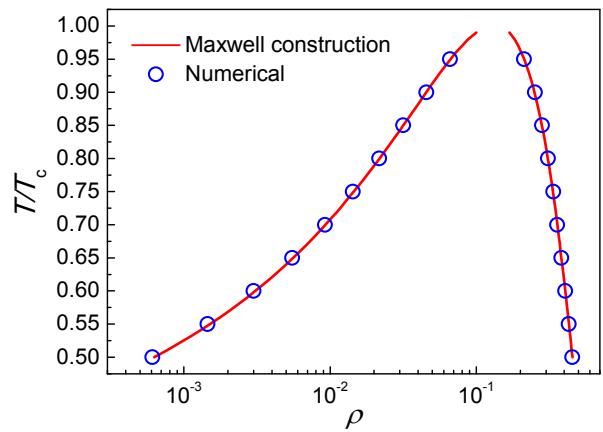

**Fig. 5** Comparison of the numerical coexistence curve predicted by the non-orthogonal MRT-LB model with the analytical coexistence curve given by the Maxwell equal-area law.

Furthermore, the non-orthogonal MRT-LB model and the orthogonal MRT-LB model [41] are compared in terms of the maximum spurious velocity through simulating static droplets. The computational domain is chosen as $N_x \times N_y \times N_z = 120 \times 120 \times 120$ and a spherical droplet of radius $r_0 = 40$ is initially located at the center of the computational domain. The reduced temperature is set to $T/T_c = 0.6$, which corresponds to the coexistence densities $\rho_L \approx 0.406$ and $\rho_V \approx 0.00308$, and the density field is initialized as follows:

$$\rho(x,y,z) = \frac{\rho_L + \rho_V}{2} - \frac{\rho_L - \rho_V}{2} \tanh\left[\frac{2(r-r_0)}{W}\right], \qquad (59)$$

where $W = 5$ is the initial interface thickness and $r = \sqrt{(x-x_0)^2 + (y-y_0)^2 + (z-z_0)^2}$, in which $(x_0, y_0, z_0)$ is the center of the computational domain. The parameter $a$ in the C-S equation of state is taken as $a = 0.25$ for the reduced temperature $T/T_c = 0.6$, which yields an interface thickness around five lattices [39]. The constant $\sigma$ in Eq. (16) is chosen as 0.105, which is different from the value of $\sigma$ for flat interfaces because the Maxwell equal-area law is theoretically established for the cases in which the pressure of liquid phase is equal to that of vapor phase, while the pressure difference across a curved interface is non-zero according to the Laplace law. An investigation of this issue can be found in Ref. [37].

**Table II**. Comparison of the numerical coexistence densities obtained by the non-orthogonal and orthogonal MRT-LB models at $T/T_c = 0.6$. The coexistence densities given by the Maxwell equal-area law are $\rho_L \approx 0.406$ and $\rho_V \approx 0.00308$.

| Model | $v = 0.01$ | | $v = 0.05$ | | $v = 0.1$ | | $v = 0.15$ | |
|---|---|---|---|---|---|---|---|---|
| | $\rho_L$ | $\rho_V$ | $\rho_L$ | $\rho_V$ | $\rho_L$ | $\rho_V$ | $\rho_L$ | $\rho_V$ |
| Non-orthogonal | 0.408 | 0.00304 | 0.408 | 0.00303 | 0.408 | 0.00300 | 0.408 | 0.00298 |
| Orthogonal | 0.408 | 0.00313 | 0.408 | 0.00319 | 0.408 | 0.00321 | 0.408 | 0.00323 |

The numerical coexistence densities predicted by the non-orthogonal MRT-LB model and the orthogonal MRT-LB model at $T/T_c = 0.6$ are shown in Table II with the kinematic viscosity varying from $v = 0.01$ to $0.15$ ($s_v^{-1}$ changing from $0.53$ to $0.95$). The table shows that the results of the non-orthogonal and orthogonal MRT-LB models are basically in good agreement with the coexistence densities given by the Maxwell equal-area law. Meanwhile, it can also be observed that there are some slight differences between the two models in the vapor density and these differences are larger than those in Table I, which may be attributed to the fact that the present test has a larger density ratio. The maximum spurious velocities given by the non-orthogonal MRT-LB model and the orthogonal MRT-LB model at $T/T_c = 0.6$ are plotted against $s_v^{-1}$ in Fig. 6, from which we can see that there are no significant differences between the results of the non-orthogonal MRTL-LB model and those of the orthogonal MRT-LB model. For both models, it can be seen that the maximum spurious velocity is smaller than $0.005$ in the cases of $s_v^{-1} \geq 0.65$, but increases to about $0.01$ when $s_v^{-1} = 0.6$, and further increases to about $0.04$ when $s_v^{-1} = 0.53$.

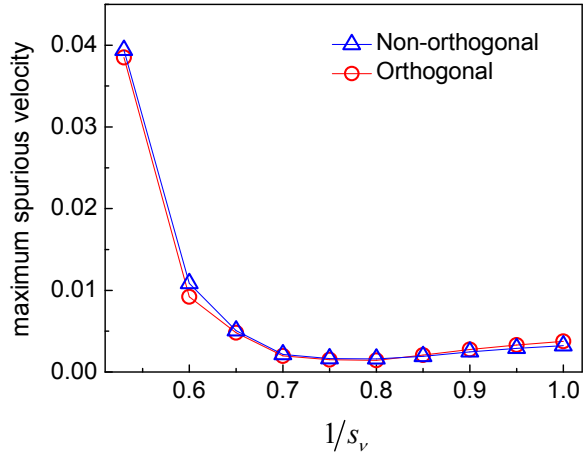

**Fig. 6** Comparison of the maximum spurious velocities given by the non-orthogonal and orthogonal MRT-LB models at $T/T_c = 0.6$.

For the results displayed in Fig. 6, the same $s_v$ is applied in the whole computational domain, which means that the vapor kinematic viscosity is equal to the liquid kinematic viscosity, namely $\nu_V = \nu_L = \nu$, and hence the ratio $\mu_L/\mu_V$ is the same as the density ratio, i.e., $\mu_L/\mu_V \equiv (\rho_L/\rho_V)/(\nu_V/\nu_L) = \rho_L/\rho_V$. In the literature [39,48] it has been shown that, besides widening the interface and using high-order isotropic gradient operator to calculate the interaction force, increasing the ratio $\nu_V/\nu_L$ can also reduce the spurious velocities in the cases of large density ratios. For example, when the liquid kinematic viscosity is taken as $\nu_L = 0.01$ ($s_v^{-1} = 0.53$ for the liquid phase), we find that the maximum spurious velocity can be reduced from about $0.04$ to about $0.006$ when the ratio $v_V/v_L$ increases from 1 to 10.

### E. Droplet impingement on a flat surface

Finally, we consider a dynamic test, the impingement of a droplet with an initial velocity on a flat surface, so as to validate the capability of the three-dimensional non-orthogonal MRT-LB model for simulating multiphase flows at large density ratios. Impingement of droplets on solid surfaces is a very important phenomenon in many engineering applications, ranging from ink-jet printing to spray cooling. In our simulations, the computational domain is taken as $N_x \times N_y \times N_z = 300 \times 300 \times 150$. Initially, a spherical droplet of diameter $D_0 = 100$ is placed on the center of the bottom surface. The initial velocity of the droplet is given by $\mathbf{u}_0 = (u_x, u_y, u_z) = (0, 0, -U_0)$, in which $U_0 = 0.075$. The no-slip boundary condition is employed at the solid surface and the periodic condition is applied in the *x* and *y* directions. The droplet dynamics is characterized by the following two non-dimensional parameters:

$$\text{We} = \frac{\rho_L D_0 U_0^2}{\vartheta}, \quad \text{Re} = \frac{\rho_L U_0 D_0}{\mu_L}, \tag{60}$$

where We and Re are the Weber number and the Reynolds number, respectively. Besides, another non-dimensional parameter can also be found in some studies, i.e., the Ohnesorge number $\text{Oh} = \mu_L / \sqrt{\rho_L \vartheta D_0}$, which is related to the Weber number and the Reynolds number via $\text{Oh} = \sqrt{\text{We}}/\text{Re}$.

In this test, the density ratio is chosen as $\rho_L/\rho_V = 800$ and the dynamic viscosity ratio is set to $\mu_L/\mu_V = 50$ (under normal temperature and atmospheric pressure, the water/air density ratio is around 830 and the corresponding dynamic viscosity ratio is about 56). A piecewise linear equation of state [37] is employed in the square-root-form pseudopotential. The surface tension $\vartheta$ is evaluated via the Laplace law and the static contact angle is taken as $\theta \approx 60°$. The Reynolds number in our simulations varies from $\text{Re} = 80$ to $\text{Re} = 1000$. Figure 7 displays some snapshots of the droplet impingement process at $\text{Re} = 1000$ and $\text{We} \approx 36$. Immediately after the impingement, it can be seen that the shape of the droplet resembles a truncated sphere ($t = 500\delta_t$). Subsequently, a lamella is formed as the liquid moves radially ($t = 1000\delta_t$). The lamella continues to grow radially and its thickness decreases ($t = 2500\delta_t$). After reaching the maximum spreading diameter, the lamella begins to retract because of the surface tension. All of these observations agree well with those reported in the previous experimental and numerical studies [41,53-55].

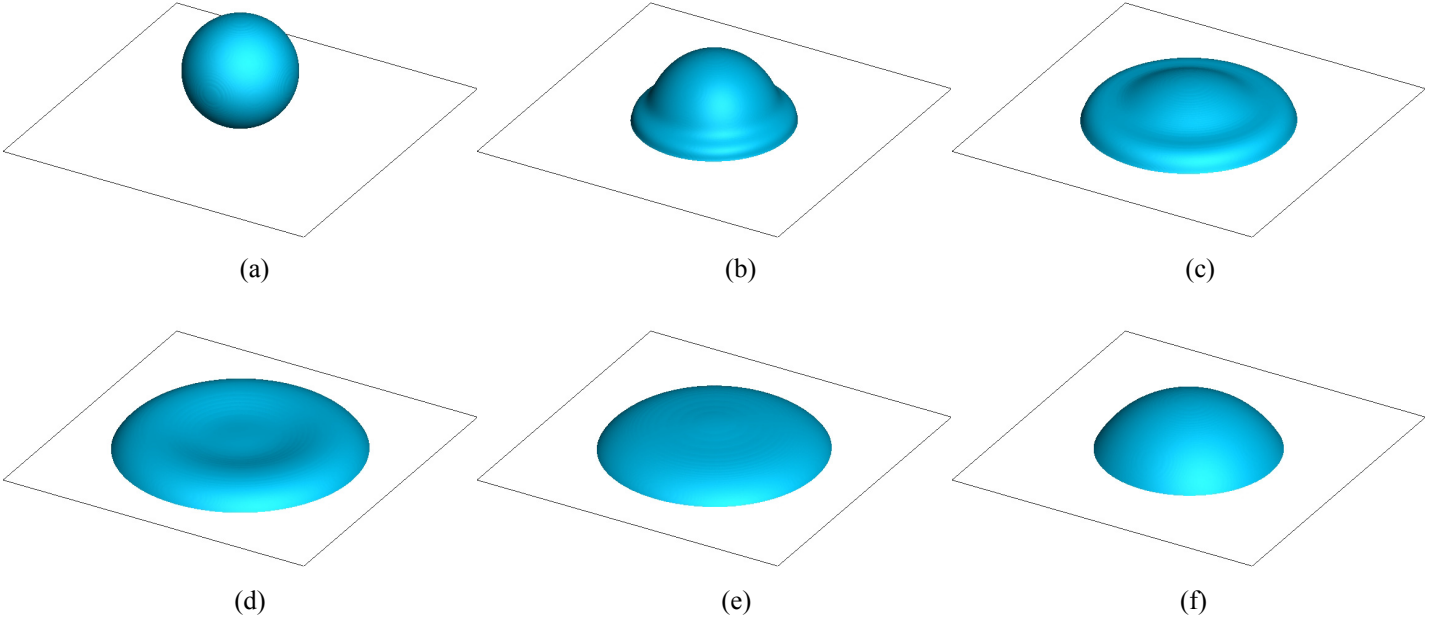

|     |     |     |
| :-: | :-: | :-: |
| (a) | (b) | (c) |
| (d) | (e) | (f) |

**Fig. 7** Snapshots of droplet impingement on a flat surface at $\rho_L/\rho_V = 800$, $\text{Re} = 1000$, and $\text{We} \approx 36$. (a) $t = 0$, (b) $t = 500\delta_t$, (c) $t = 1000\delta_t$, (d) $t = 2500\delta_t$, (e) $t = 5000\delta_t$, and (f) $t = 20000\delta_t$.

In the literature, the maximum spreading factor $D_{\max}/D_0$ is usually employed to quantify the numerical results [41,55]. In Ref. [53], Asai *et al.* established a correlation formula for the maximum spreading factor according to their experimental data: $D_{\max}/D_0 = 1 + 0.48 \text{We}^{0.5} \exp\left(-1.48 \text{We}^{0.22} \text{Re}^{-0.21}\right)$. Scheller and Bousfield [54] have also proposed a correlation formula by plotting their experimental data against $\text{Oh}\,\text{Re}^2 \equiv \sqrt{\text{We}}\,\text{Re}$. A comparison of the maximum spreading factor between the correlation formula of Asai *et al.*, the experimental data of Scheller and Bousfield, and the present simulation results is provided in Fig. 8, where the maximum spreading factor $D_{\max}/D_0$ is plotted against $\text{Oh}\,\text{Re}^2 = \sqrt{\text{We}}\,\text{Re}$. The figure shows that our numerical results are in good agreement with the experimental correlation/data in the previous studies, which demonstrates the capability of the three-dimensional non-orthogonal MRT-LB for simulating multiphase flows at large density ratios.

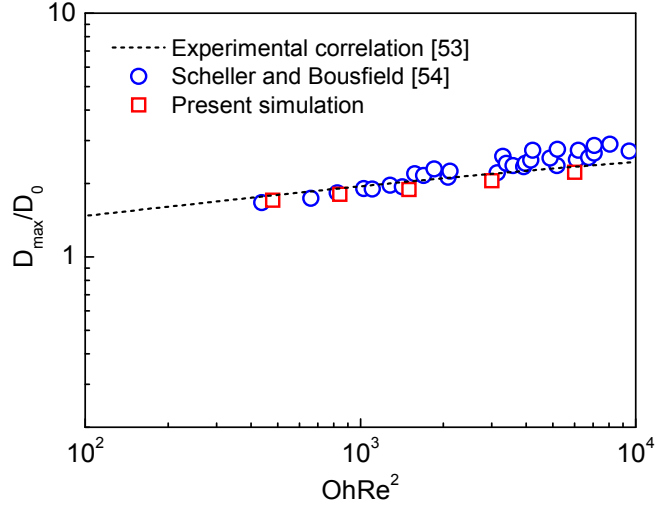

**Fig. 8** Comparison of the maximum spreading factor between the present simulation results, the experimental correlation in Ref. [53], and the experimental data in Ref. [54].

## V. Summary

A theoretical and numerical study has been performed to investigate the capability and efficiency of a three-dimensional non-orthogonal MRT-LB method for simulating multiphase flows. The model is developed based on the D3Q19 lattice with a non-orthogonal MRT collision operator, which is devised from a set of non-orthogonal basis vectors for the D3Q19 lattice. Using the non-orthogonal MRT collision operator, the transformation matrix $\mathbf{M}$ and its inverse matrix $\mathbf{M}^{-1}$ are much simpler than those of the usual orthogonal MRT collision operator. Through the Chapman-Enskog analysis, it has been demonstrated that the three-dimensional non-orthogonal MRT-LB model can correctly recover the Navier-Stokes equations in the low Mach number limit. Numerical investigations have been carried out based on the pseudopotential multiphase LB approach. Both the exponential-form pseudopotential and the square-root-form pseudopotential have been considered in our simulations. Numerical comparisons show that the non-orthogonal MRT-LB model retains the numerical accuracy when simplifying the implementation, and can serve as an alternative to the usual orthogonal MRT-LB model.

## Acknowledgments

This work was supported by the National Natural Science Foundation of China (No. 51506227), the Foundation for the Author of National Excellent Doctoral Dissertation of China (No. 201439), and the UK Consortium on Mesoscale Engineering Sciences (UKCOMES) under the UK Engineering and Physical Sciences Research Council Grant No. EP/R029598/1. Part of this work was done when the first author was working at the Los Alamos National Laboratory supported by the Director's Funded Fellowship.

## Appendix A: The inverse matrix of the non-orthogonal transformation matrix

According to the non-orthogonal transformation matrix defined by Eq. (12), its inverse matrix is given by

$$\mathbf{M}^{-1} = \begin{bmatrix}
1 & 0 & 0 & 0 & -1 & 0 & 0 & 0 & 0 & 0 & 0 & 0 & 0 & 0 & 0 & 0 & 1 & 1 & 1 \\
0 & 0.5 & 0 & 0 & 1/6 & 1/6 & 0 & 0 & 0 & 0 & 0 & -0.5 & 0 & -0.5 & 0 & 0 & -0.5 & -0.5 & 0 \\
0 & -0.5 & 0 & 0 & 1/6 & 1/6 & 0 & 0 & 0 & 0 & 0 & 0.5 & 0 & 0.5 & 0 & 0 & -0.5 & -0.5 & 0 \\
0 & 0 & 0.5 & 0 & 1/6 & -1/12 & 0.25 & 0 & 0 & 0 & -0.5 & 0 & 0 & 0 & 0 & -0.5 & -0.5 & 0 & -0.5 \\
0 & 0 & -0.5 & 0 & 1/6 & -1/12 & 0.25 & 0 & 0 & 0 & 0.5 & 0 & 0 & 0 & 0 & 0.5 & -0.5 & 0 & -0.5 \\
0 & 0 & 0 & 0.5 & 1/6 & -1/12 & -0.25 & 0 & 0 & 0 & 0 & 0 & -0.5 & 0 & -0.5 & 0 & 0 & -0.5 & -0.5 \\
0 & 0 & 0 & -0.5 & 1/6 & -1/12 & -0.25 & 0 & 0 & 0 & 0 & 0 & 0.5 & 0 & 0.5 & 0 & 0 & -0.5 & -0.5 \\
0 & 0 & 0 & 0 & 0 & 0 & 0 & 0.25 & 0 & 0 & 0.25 & 0.25 & 0 & 0 & 0 & 0 & 0.25 & 0 & 0 \\
0 & 0 & 0 & 0 & 0 & 0 & 0 & 0.25 & 0 & 0 & -0.25 & -0.25 & 0 & 0 & 0 & 0 & 0.25 & 0 & 0 \\
0 & 0 & 0 & 0 & 0 & 0 & 0 & -0.25 & 0 & 0 & -0.25 & 0.25 & 0 & 0 & 0 & 0 & 0.25 & 0 & 0 \\
0 & 0 & 0 & 0 & 0 & 0 & 0 & -0.25 & 0 & 0 & 0.25 & -0.25 & 0 & 0 & 0 & 0 & 0.25 & 0 & 0 \\
0 & 0 & 0 & 0 & 0 & 0 & 0 & 0 & 0.25 & 0 & 0 & 0 & 0.25 & 0.25 & 0 & 0 & 0 & 0.25 & 0 \\
0 & 0 & 0 & 0 & 0 & 0 & 0 & 0 & 0.25 & 0 & 0 & 0 & -0.25 & -0.25 & 0 & 0 & 0 & 0.25 & 0 \\
0 & 0 & 0 & 0 & 0 & 0 & 0 & -0.25 & 0 & 0 & 0 & 0 & -0.25 & 0.25 & 0 & 0 & 0 & 0.25 & 0 \\
0 & 0 & 0 & 0 & 0 & 0 & 0 & -0.25 & 0 & 0 & 0 & 0 & 0.25 & -0.25 & 0 & 0 & 0 & 0.25 & 0 \\
0 & 0 & 0 & 0 & 0 & 0 & 0 & 0 & 0 & 0.25 & 0 & 0 & 0 & 0 & 0.25 & 0.25 & 0 & 0 & 0.25 \\
0 & 0 & 0 & 0 & 0 & 0 & 0 & 0 & 0 & 0.25 & 0 & 0 & 0 & 0 & -0.25 & -0.25 & 0 & 0 & 0.25 \\
0 & 0 & 0 & 0 & 0 & 0 & 0 & 0 & 0 & -0.25 & 0 & 0 & 0 & 0 & -0.25 & 0.25 & 0 & 0 & 0.25 \\
0 & 0 & 0 & 0 & 0 & 0 & 0 & 0 & 0 & -0.25 & 0 & 0 & 0 & 0 & 0.25 & -0.25 & 0 & 0 & 0.25
\end{bmatrix}.$$

## Appendix B: Comparison of the non-orthogonal and orthogonal models in terms of the computational cost and the Mach number effect.

In this Appendix, the three-dimensional non-orthogonal MRT-LB model is compared with the usual orthogonal MRT-LB model in terms of the computational cost and the Mach number effect through modeling the three-dimensional lid-driven cavity flow [56]. The grid system is taken as $N_x \times N_y \times N_z = L \times L \times L$ and the driving lid is placed at $y = L$, moving along the direction of $x$-axis with a speed $U$. First, the computation cost is compared. Three different values of $L$ are considered ($L = 60$, 80, and 110) and the program runs on a desktop machine equipped by an Intel(R) Core(TM) i7-4790 CPU-3.60 GHz. The CPU time is measured after 5000 iterations and the results are shown in Table III. From the table it can be seen that the non-orthogonal MRT-LB model is about 15% faster than the orthogonal MRT-LB model.

**Table III.** Comparison of the computational cost between the orthogonal and non-orthogonal MRT-LB models. The CPU time (s) is measured after accomplishing 5000 iterations.

| $L$ | $t_{\text{orth}}$ | $t_{\text{non-orth}}$ | $t_{\text{orth}}/t_{\text{non-orth}} - 1$ |
|---|---|---|---|
| 60 | 355.48 | 311.83 | 14.0% |
| 80 | 833.72 | 730.18 | 14.2% |
| 110 | 2208.88 | 1901.62 | 16.2% |

Furthermore, the steady numerical results of the two models for the three-dimensional lid-driven cavity flow are compared. To investigate the effect of the Mach number, three cases are considered, i.e., $U = 0.1$, 0.3, and 0.5, which correspond to the Mach numbers $\text{Ma} = U/c_s \approx 0.058$, 0.173, and 0.289, respectively. The Reynolds number is chosen as $\text{Re} = UL/\nu = 400$ and $L$ is set to 80. Comparisons of the transversal velocity $u_y(x)$ at $y/L = 0.5$ and $z/L = 0.5$ and the horizontal velocity $u_x(y)$ at $x/L = 0.5$ and $z/L = 0.5$ are made in Fig. 9, in which the driving speed is $U = 0.1$ and the Mach number is $\text{Ma} \approx 0.058$, fulfilling the low Mach number limit. From the figure we can see that the numerical results of the non-orthogonal and orthogonal MRT-LB models are both in good agreement with the results reported in the study of Mei *et al.* [56], and there are no visible differences between the results of the two models.

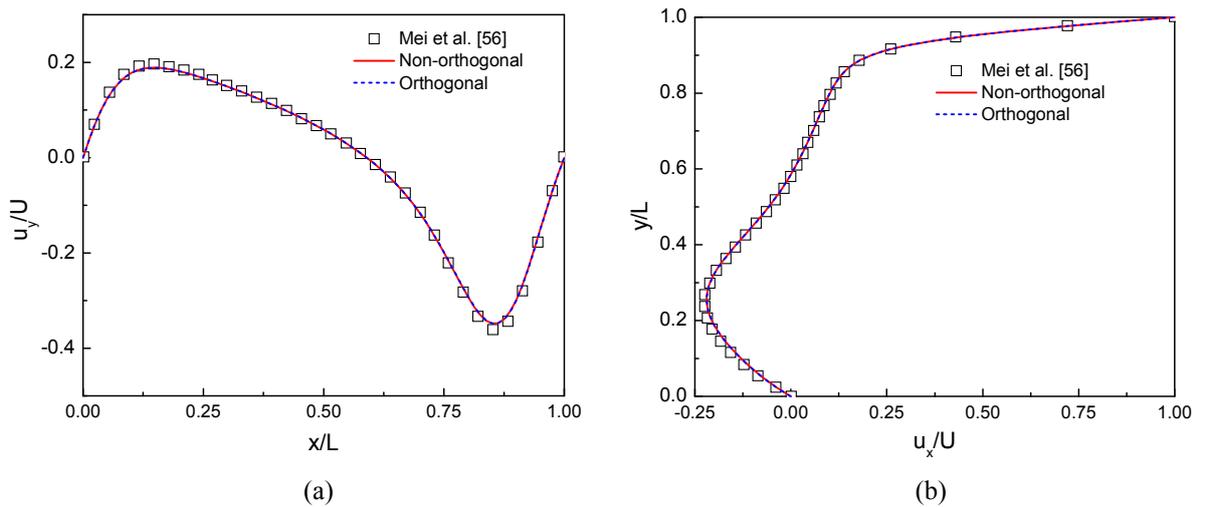

**Fig. 9** Comparison of the numerical results obtained by the non-orthogonal and orthogonal MRT models with the results reported in Ref. [56]. (a) The transversal velocity $u_y(x)$ at $y/L = 0.5$ and $z/L = 0.5$. (b) The horizontal velocity $u_x(y)$ at $x/L = 0.5$ and $z/L = 0.5$. The driving speed is $U = 0.1$.

Figure 10 displays the influence of the Mach number on the numerical results of the non-orthogonal and orthogonal MRT-LB models. From the figure it can be seen that the deviations between the numerical results of the two MRT-LB models and the results reported in Ref. [56] are getting larger when the Mach number ($\text{Ma} = U/c_s$) increases, confirming that the third-order velocity terms neglected in Eq. (50) gradually have an important influence with the increase of the Mach number. Meanwhile, in Fig. 10 there are no significant differences between the results of the non-orthogonal MRT-LB model and those of the orthogonal MRT-LB model, which implies that the two models behave the same in terms of the Mach number effect.

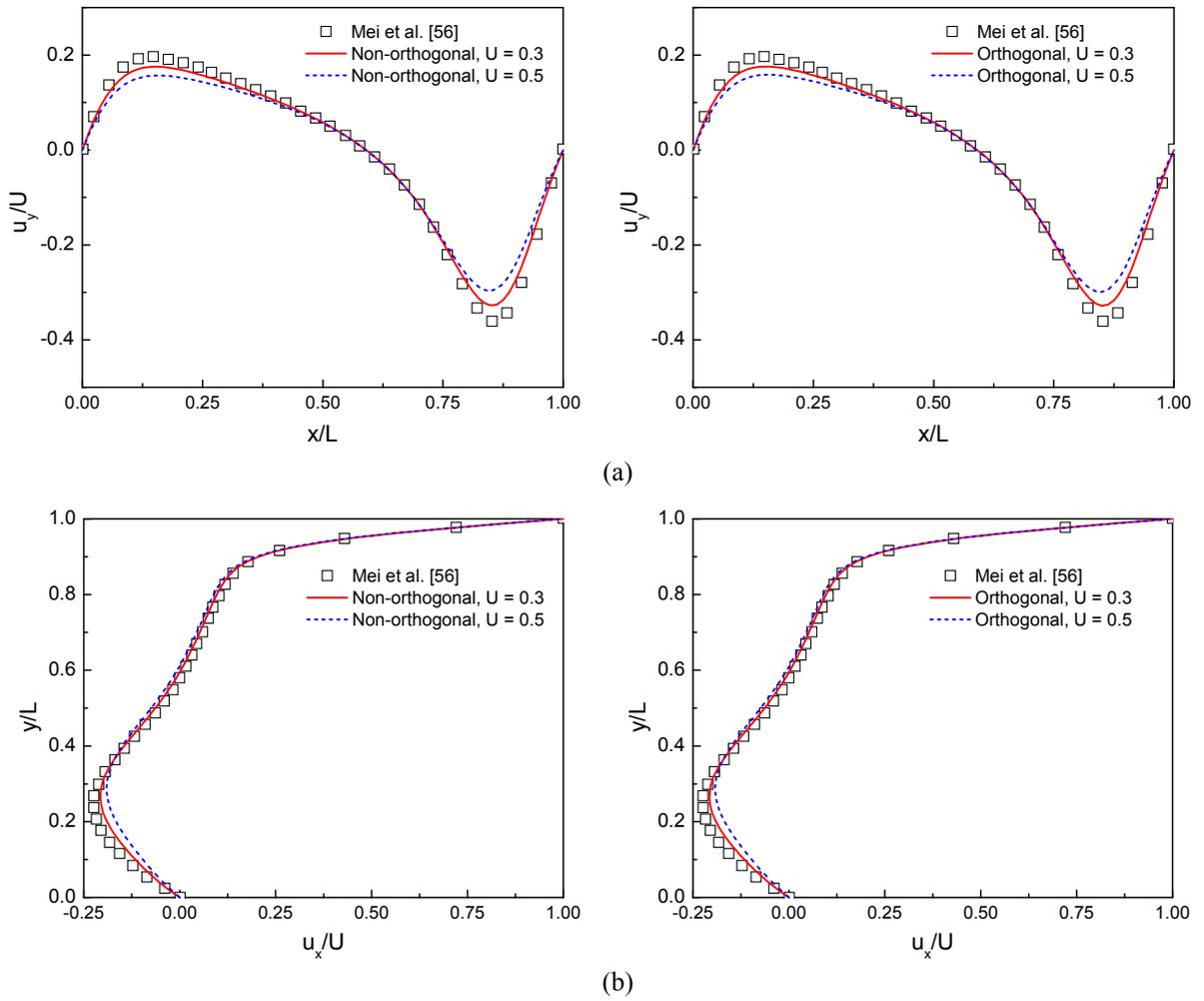

**Fig. 10** Effect of the Mach number. The numerical results are obtained by the non-orthogonal MRT-LB model (left) and the orthogonal MRT-LB model (right) with $U = 0.3$ and $0.5$, which correspond to the Mach numbers $\text{Ma} = U/c_s \approx 0.173$ and $0.289$, respectively. (a) The transversal velocity $u_y(x)$ at $y/L = 0.5$ and $z/L = 0.5$. (b) The horizontal velocity $u_x(y)$ at $x/L = 0.5$ and $z/L = 0.5$.